\documentstyle[aps,pre]{revtex}

\newcommand{\beq}{\begin{equation}}
\newcommand{\beqn}{\begin{displaymath}}
\newcommand{\zen}{\end{equation}}
\newcommand{\zenn}{\end{displaymath}}

\newcommand{\pom}{\mbox{I$\!$P}}
\def\pnot{\mbox{${\not{\hbox{\kern-3.0pt$p$}}}$}}
\def\qnot{\mbox{${\not{\hbox{\kern-2.0pt$q$}}}$}}
\def\enot{\mbox{${\not{\hbox{\kern-2.0pt$e$}}}$}}
\def\knot{\mbox{${\not{\hbox{\kern-2.0pt$k$}}}$}}

\def\fun#1#2{\lower3.6pt\vbox{\baselineskip0pt\lineskip.9pt\ialign
{$\mathsurround=0pt#1\hfil##\hfil$\crcr#2\crcr\sim\crcr}}}

\begin{document}
\begin{titlepage}

\hskip 12cm 
\rightline{\vbox{\hbox{DFPD 96/TH 11}\hbox{CS-TH 2/96} \hbox{Feb. 1996}}}
\vskip 0.6cm
\centerline{\bf NON-LINEAR POMERON TRAJECTORY IN}
\centerline{\bf DIFFRACTIVE DEEP INELASTIC SCATTERING$^{~\diamond}$}
\vskip 1.0cm
\centerline{  R. Fiore$^{a\dagger}$, L. L. Jenkovszky$^{b\ddagger}$, F. 
Paccanoni$^{c\ast}$}
\vskip .5cm
\centerline{$^{a}$ \sl  Dipartimento di Fisica, Universit\`a della Calabria,}
\centerline{\sl Istituto Nazionale di Fisica Nucleare, Gruppo collegato di 
Cosenza}
\centerline{\sl Arcavacata di Rende, I-87030 Cosenza, Italy}
\vskip .5cm
\centerline{$^{b}$ \sl  Bogoliubov Institute for Theoretical Physics,}
\centerline{\sl Academy of Sciences of the Ukrain}
\centerline{\sl 252143 Kiev, Ukrain}
\vskip .5cm
\centerline{$^{c}$ \sl  Dipartimento di Fisica, Universit\`a di Padova,}
\centerline{\sl Istituto Nazionale di Fisica Nucleare, Sezione di Padova}
\centerline{\sl via F. Marzolo 8, I-35131 Padova, Italy}
\vskip 1cm
\begin{abstract}
Recent experimental data on diffractive deep inelastic scattering
collected by the H1 and ZEUS Collaborations at HERA are analysed
in a model with a non-linear trajectory in the pomeron flux. The
$t$ dependence of the diffractive structure function $F_2^{D(4)}$ is
predicted. The normalization of the pomeron flux and the (weak)
$Q^2$ dependence of the pomeron structure function are revised as
well.
\end{abstract}
\vskip .5cm
\hrule
\vskip .3cm
\noindent

\noindent
$^{\diamond}${\it Work supported by the Ministero italiano
dell'Universit\`a e della Ricerca Scientifica e Tecnologica, by the EEC 
Programme ``Human Capital and Mobility", Network ``Physics at High
Energy Colliders", contract CHRX-CT93-0357 (DG 12 COMA) and by the INTAS 
Grant 93-1867}
\vfill
$\begin{array}{ll}
^{\dagger}\mbox{{\it email address:}} &
   \mbox{FIORE~@CS.INFN.IT}
\end{array}
$

$ \begin{array}{ll}
^{\ddagger}\mbox{{\it email address:}} &
 \mbox{JENK~@GLUK.APC.ORG}
\end{array}
$

$ \begin{array}{ll}
^{\ast}\mbox{{\it email address:}} &
   \mbox{PACCANONI~@PADOVA.INFN.IT}
\end{array}
$
\vfill
\end{titlepage}
\eject
\textheight 210mm
\topmargin 2mm
\baselineskip=24pt
{\bf 1. Introduction}

Diffractive deep inelastic scattering, or ''hard diffraction'', is
a subclass of semi-inclusive deep inelastic reactions in which a
virtual point-like particle (quark or lepton) is assumed to interact
directly with the pomeron emitted by the proton target. This may
happen in a special kinematical configuration when the hit proton
continues its motion nearly in the forward direction.
\vskip 0.3cm
There are several reasons why this class of reactions is receiving
so much attention. The first and most obvious one comes from the 
prospect to study the internal structure of the pomeron, a
hypothetical quasi-particle with the quantum numbers of vacuum,
responsible for diffraction in the Regge pole model. Although the
composite nature of the pomeron, within the context of the quark
model and QCD, was never a question for the theorists, the
possibility of its experimental verification in deep inelastic
scattering was first formulated in Ref. ~\cite{SCH}. In subsequent 
experiments of the UA8 Collaboration ~\cite{UA8}, 2-jet events were
reported as evidence of the partonic structure of the pomeron.
\vskip 0.3cm
The study of the so-called ''large rapidity gap'' events at HERA
initiated by the ZEUS Collaboration ~\cite{ZEUS0}, revitalized the
subject. HERA produced a large number of high-precision data in
a wide kinematical range of $x$ and $Q^2$, published recently by
the H1 ~\cite{H1} and ZEUS ~\cite{ZEUS1} Collaborations. A new
and important development in the subject is the experimental study
of the $t$ dependence of the cross-sections, ignored until 
recently.
\vskip 0.3cm
Theoretical studies of the diffractive deep inelastic scattering 
(hard diffraction) were pioneered by the papers of Donnachie and
Landshoff ~\cite{DL}, long before the HERA experimentations.
Now there is a large literature (partially listed in Refs. 
~\cite{BERG,PRYTZ,BUCH,ALL0,ALL,GOUL,CF},
we apologize to those not included because of the limited scope of
the present paper) dealing with various aspects of the phenomenon.
In our opinion, the present state of the subject can be summarized
as follows.
\vskip 0.3cm
There is nothing surprising in the very existence of the large 
rapidity gap events (occurring e.g. also in the $p\bar{p}$
diffraction dissociation)  provided
diffraction, the Regge pole model and properties of the S matrix
do not change essentially when one of the external particles
goes off shell. The last assumption is the most delicate one in
the whole subject, and we shell come back to it.
\vskip 0.3 cm
The next question is whether the appearance of a rapidity gap,
within which secondary particles are not produced, is a manifestation
of diffraction, and if so: is it a pomeron exchange, or can it be
simulated by an alternative mechanism ?. The first answer is ''yes''
- to the extent other features of diffraction, namely typical
$\xi$ (or $x_{\pom}$) and $t$ dependence will be confirmed.
\vskip 0.3cm
The $\xi$ dependence has been recently measured and found to be
$\approx \xi^{-a}$ with the values $a=1.30\pm 0.08(stat)
\begin{array}{c} +0.08 \\ -0.14 \end{array}(sys)$ as measured by ZEUS
~\cite{ZEUS1} and $a=1.19\pm 0.06(stat)\pm 0.07(sys)$ by H1 ~\cite{H1}.
These two values are compatible and both agree with the pomeron
intercept ~\cite{DL2} most reliably extracted from the $p-p$
and $p-\bar{p}$ total cross section. The $t$ dependence is now being 
measured and we discuss it below.
\vskip 0.3cm
Whether diffraction is mediated by a pomeron (pole) or not is a
matter of convention. A factorizable vacuum Regge pole exchange
with the trajectory whose intercept is slightly beyond 1, $\alpha(0)=
1+\epsilon$, is known to be the adequate mechanism of diffraction, at 
least in hadronic reactions. Whether and how is the formalism 
affected by off mass shell effects is an open question. In any
case the use of the pomeron exchange is more justified in diffractive
DIS, where at least the pomeron interaction is directly related
to hadronic diffraction, than generally in a (small-$x$) DIS,
related to the absorptive part of the elastic $\gamma^*-p$ 
scattering only for $t=0$.
\vskip 0.3cm
The large $Q^2$ flow along the photon line was an argument for
many authors to use perturbative QCD calculations for a single-
~\cite{BUCH} or two-gluon (pomeron) ~\cite{ALL0}
 exchange in diffractive DIS. In our
opinion, the use of perturbative QCD may be justified only in
resolving the pomeron structure (QCD evaluation of the photon
pomeron vertex), while the exchanged object, the pomeron propagator,
is essentially nonperturbative and it is the same, whatever is the
photon virtuality. A more detailed discussion concerning the uniqueness 
of the pomeron in elastic and deep inelastic scattering may be found
in Ref. ~\cite{BGJPP}.
\vskip 0.3cm
The above construction relies very much on the hypotheses of
factorization of the diffractive structure function $F_2^{D(4)}$
into a product of the pomeron flux and the pomeron structure
function. Small deviations from factorization, e.g. due to 
multipomeron exchanges, initial and final state interactions etc.
are admissible and certainly should be accounted for in future
calculations. At the moment, according to the recent measurements
at HERA ~\cite{H1,ZEUS1}, factorization is confirmed at
the present level of accuracy. 
\vskip 0.3cm
The ultimate goal of the experiments at HERA and of their theoretical
interpretation is to answer the question: what is the pomeron structure
or, in other words, to find the right parametrization for the 
pomeron structure function. Because of the complexity of the problem and 
the existing uncertainties, both on the experimental and theoretical
side, the final result is still ambiguous and sensitive to the
input.
\vskip 0.3cm
With the present paper, we try to contribute by one more step in the
clarification of the remaining uncertainties. First of all in Sec.{\bf 2} we
introduce notations and define the kinematics.  Next in Sec.{\bf 3} we
discuss the pomeron flux and its (re)normalization. The 
pomeron structure function and the comparison with the experimental
data are developed in Sec.{\bf 4} while Sec.{\bf 5} is devoted
to a study of the $t$ dependence of the distribution function.
We use the earlier experience based on the description of 
hadron diffraction with a non-linear pomeron trajectory ~\cite{DJL}
to predict the $t$ dependence of the diffractive DIS.
  
\vskip 0.5cm

{\bf 2. Notations and kinematics.}
 
The notations and kinematics for the process
\beq
 e^-(k)+p(p) \longrightarrow e^-(k')+p(p')+X(p_X)
\label{z1}
\zen
are standard. The fourmomenta of the virtual photon and of the
exchanged reggeon are $q=k-k'$ and $r=p-p'$, respectively. Besides
the usual DIS variables $Q^2=-q^2, x=Q^2/(2p\cdot q),$  and
$W^2=(p+q)^2$ the new variables
\beq
\beta=\frac{Q^2}{2r\cdot q},\hspace{7mm} \xi=\frac{r\cdot q}{p\cdot q}
=\frac{x}{\beta}
\label{z2}
\zen
are introduced. Sometimes $\xi$ is called also $x_{\pom}$.
\vskip 0.3cm
Let $M_X$ be the invariant mass of the hadronic system $X$, $p_X^2=
M_X^2$, then the fourmomentum transfer squared, $t=r^2$, ranges
between the extreme limits
\beq
t_{\mp}=\left(\frac{M_X^2+Q^2}{2W}\right)^2-\left\{\left[\left(
\frac{W^2-Q^2-m_p^2}{2W}\right)^2+Q^2\right]^{1/2}\mp\left[
\left(\frac{W^2+M_X^2-m_p^2}{2W}\right)^2-M_X^2\right]^{1/2}
\right\}^2
\label{z3}
\zen
or
\beqn
t_-\sim -m_p^2x^2\left(1+\frac{M_X^2}{Q^2}\right)^2\sim -m_p^2\xi^2
\zenn
while $t_+\sim -W^2$ is a formal limit since $|t|$ is less than
$\sim 7\,GeV^2$ for the measurement in ~\cite{H1}. Experimentally,
the $t$ distribution has never been measured until recently because
of the existing difficulties in identifying the proton hit by the
photon but continuing its motion in the nearly forward direction.
\vskip 0.3cm
On the other hand it is well known from the hadronic physics that
diffraction is typical of the domain of about $|t|\leq 2\,GeV^2$.
In this region, the elastic differential cross section is known to
decrease almost exponentially up to about $|t|\simeq 1\,GeV^2$,
followed by a dip-bump structure between $1$ and $2\,GeV^2$.
The latter is an important feature of high energy diffraction. The
deviation from the exponential behaviour of the cone may be taken 
into account by a nonlinear trajectory, a simple representative
example of which is ~\cite{DJL}
\beq
\alpha(t)=\alpha_0+\alpha_1t-\alpha_2\ln(1-\alpha_3t)
\label{z4}
\zen
where $\alpha_0=1+\epsilon$ and $\alpha_i (i=1,2,3)$ are parameters.
\vskip 0.3cm
The diffractive structure function is defined from the cross section
as in ~\cite{PRYTZ} and the factorization hypothesis can be 
written as
\beq
F_2^{D(4)}(x,Q^2,t,\xi)=F_{\pom/p}(\xi,t) G_{q/\pom}(\beta,Q^2).
\label{z5}
\zen
$F_{\pom/p}(\xi,t)$ and $G_{q/\pom}(\beta,Q^2)$ are, respectively, the
pomeron flux and the pomeron structure function, to be introduced
in the next two sections.
We quote also the upper and lower kinematical limits for $\xi$
typical of the H1 and ZEUS data:
\vskip 0.4cm
\begin{center}
\begin{tabular}{|c|l|l|}  \hline\hline
     &  H1  &  ZEUS \\ \hline
$\xi_H$ & 0.05 & 0.01 \\ \hline
$\xi_L$ & $3\times 10^{-4}$ & $6.3\times 10^{-4}$ \\ \hline\hline
\end{tabular}
\end{center}

\vskip 0.7cm

{\bf 3. The pomeron flux and its (re)normalization}

As in Ref. ~\cite{FJP} we fix the form of the scattering amplitude
following the duality prescription. In dual models the dependence
on the Mandelstam variable $t$ enters the amplitude only through the
trajectory $\alpha(t)$. Hence in the Regge limit, $s\rightarrow
\infty$ at fixed $t$,
\beq
A(s,t)=e^{B(s) \alpha(t)}~,
\label{z6}
\zen
where $B(s)=B_{el}+\ln(s/s_0)-i\pi/2$.
\vskip 0.3cm
We have already discussed ~\cite{FJP}
the relevance of a nonlinear trajectory that
extrapolates between ''soft'' and ''hard'' scattering. The interest
for a possible flattening of the pomeron trajectory increased in
view of the recent experimental result of the UA8 Collaboration
~\cite{SCL2}. In order to make clear the effect of a nonlinear 
$\alpha(t)$ we will consider Eq. (\ref{z4}) in the two limiting cases
\beq
a)\,\alpha_2=0:\ \alpha(t)=\alpha(0)+\alpha' t~,
\label{z7}
\zen
\beq
b)\,\alpha_1=0, \alpha_2=\alpha_3:\ \alpha(t)=\alpha(0)-\gamma
\ln(1-\gamma t)~.
\label{z8}
\zen
The first instance has been studied in Ref. ~\cite{FJP} for different
choices of $\alpha(0)$ and $\alpha'=0.25$. In case $b)$ we will 
choose $\gamma=0.5$ in order to reproduce the same slope near $t=0$
and to deal with a trajectory close to the optimal solution of 
Ref. ~\cite{DJL}. 
\vskip 0.3cm
The constant term in $B(s), B_{el},$ has been determined from a fit to
the elastic $p-p$ differential cross section at ISR ~\cite{SCHI}.
For the pomeron flux
\beq
F_{\pom/p}=C e^{2(B-\ln\xi)\alpha(t)}\cdot\xi
\label{z9}
\zen
we get
\beqn
B=B_{diff}=B_{el}/2\simeq 7.0.
\zenn
The main hypothesis here is that only the pomeron trajectory contributes
to the single diffractive cross section, a condition supported from 
the experimental evidence for factorization. Since the trajectory
(\ref{z8}) is an effective one, the best determination of $\alpha(0)$
comes from the overall Regge fit of total cross sections in Ref. ~\cite{DL2} 
\beqn
\alpha(0)\simeq 1.08.
\zenn
We will use in the following this value that differs slightly from the
CDF intercept $\alpha(0)=1.112\pm 0.013$ ~\cite{CDF} obtained from total
cross section data, but provides a sensible determination of the
constant in front of the flux (\ref{z9}).
\vskip 0.3cm
From ~\cite{DL2} we get the pomeron contribution to $p-p$ total cross
section and find
\beq
c\equiv C e^{2B\alpha(0)}=\frac{21.7}{16\pi} mb= 1.1087\,GeV^{-2}.
\label{z10}
\zen
From the same fit we deduce that the contribution to the cross section
of non-asymptotic Regge terms ($\rho, f,\ldots$) is rather small at
HERA, approximately $8\%$ at the lowest $W$ values. As seen in Fig. 1,
there is a marked difference in the $t$ dependence of the pomeron flux
derived from a logarithmic trajectory, curve (1), or a linear one,
curve (2). Curves (3), ~\cite{BERG} and (4), ~\cite{DL} refer to 
models for the pomeron flux approaching the limiting cases 
represented by eqs. (\ref{z7}) and (\ref{z8}).
\vskip 0.3cm
The integrated pomeron flux for the nonlinear trajectory (\ref{z8}) is
\beqn
\Phi(\xi)\equiv \int_{t_+}^{t_-}\,F_{\pom/p}(\xi,t) dt=
\zenn
\beq
\frac{c}{\gamma}\frac{(1-\gamma t_-)^{-b(\xi)}-
(1-\gamma t_+)^{-b(\xi)}}{b(\xi)} \xi^{1-2\alpha(0)}\simeq
\frac{c}{\gamma b(\xi)} \xi^{1-2\alpha(0)}
\label{z11}
\zen
where
\beqn
b(\xi)=2\gamma(B-\ln\xi)-1.
\zenn
The approximation in (\ref{z11}) is reasonable since $t_-\propto
-\xi^2$ and $|t_+|$ is rather large, $|t_+|\simeq 7\,GeV^2$ for the
measurement described in Ref. ~\cite{H1}.
\vskip 0.3cm
Now the $\xi,t$ dependence of the diffractive structure function
is completely fixed within the model. In order to set the normalization 
and to make a comparison with the experimental data we need the 
pomeron structure function.
\vskip 0.3cm
Renormalization of the pomeron flux ~\cite{GOUL} affects only the
determination of the pomeron structure function. If we require that
no more than one pomeron should be exchanged in one diffractive proton
interaction, then a bound must be imposed on the integral
\beq
\int_{\xi_L}^{\xi_H}\,\Phi(\xi) d\xi\simeq
\frac{c}{2\gamma^2}\,e^{-f(1)} \{Ei[f(\xi_L)]-Ei[f(\xi_H)]\}~,
\label{z12}
\zen
where
\beqn
f(\xi)=2(\alpha(0)-1)(B-\frac{1}{2\gamma}-\ln\xi)
\zenn
and $\xi_H$, $\xi_L$ are the upper and lower kinematical limits for
$\xi$ specified in Sec. {\bf 2}. In Eq. (\ref{z12}), $Ei(z)$ is the 
exponential integral ~\cite{BAT}, and the numerical values of the
integrated flux are $1.34$ for ZEUS ~\cite{ZEUS1} and $2.42$ for
H1 ~\cite{H1}. For a linear trajectory (\ref{z7}) these values are 
somewhat smaller: $1.23$ and $2.23$ respectively. Considering the
unitarization procedure explained in Ref. ~\cite{GOUL}, flux
renormalization should be applied also to some $(Q^2,\beta)$ bins
of the ZEUS data, in particular for small $Q^2$ and large $\beta$.
Since corrections remain within experimental errors we do not
scale down the integrated flux.

\vskip 0.5cm

{\bf 4. The pomeron structure function and comparison with experimental
data.}

As in our previous paper ~\cite{FJP} we consider a pomeron composed 
mainly of gluons. This point of view reflects the structure of the 
BFKL pomeron ~\cite{BFKL} that, at least for large momentum transfers,
relies on a sound basis. Non perturbative effects will represent a
new contribution to the aforesaid picture but should not change the
pomeron content. On its gluon content both theoretical ~\cite{CF}
and experimental ~\cite{PHOTO} constraints exist. Experimental data 
indicate that a large fraction of the pomeron momentum, up to $80\%$,
can be carried by hard gluons ~\cite{PHOTO} and that the pomeron 
structure function is approximately independent of $Q^2$
 ~\cite{H1,ZEUS1}.
\vskip 0.3cm
Since only quarks interact with the photon, quarks must be present in
the input distribution with a fraction of the pomeron momentum larger 
than $20\%$. In Ref. ~\cite{FJP} we noticed that, at large $\beta$, to the
assumed gluon distribution
\beq
\beta G(\beta,Q^2)=a(Q^2)(1-\beta)
\label{z13}
\zen
one must add the $q\bar{q}$ sea contribution. Quark loops and the 
mesonic trajectories, included in the effective trajectory (\ref{z8}),
will give rise to a quark distribution
\beq
\beta q_i(\beta,Q^2)=d(Q^2)\beta(1-\beta)
\label{z14}
\zen
for the quark $i$, whose form we borrow from Ref. ~\cite{DL}. Both
distributions contribute at the starting scale $Q_0^2= 5\,GeV^2$ and,
due to the presence of quarks, the previous estimate of the 
evolution ~\cite{FJP} does not hold now.
\vskip 0.3cm
We can keep however the calculation simple and transparent if we
neglect gluon recombination effects ~\cite{PRYTZ} and limit 
ourselves to the low-$Q^2$ region, $Q^2\leq 40\,GeV^2.$ This
condition will not destroy the predictive power of the result since
''there is no evidence for any substantial $Q^2$ dependence of
$\tilde{F}_2^D$'' ~\cite{H1}, a function proportional to the pomeron
structure function.
\vskip 0.3cm
We use a recursive method for solving the massless inhomogeneous
Altarelli-Parisi equations. This method is based on the power expansion
of the solution in the parameter $s$:
\beqn
s=\ln\left(\frac{\ln(Q^2/\Lambda^2)}{\ln(Q_0^2/\Lambda^2)}\right)
\zenn
and, since it has been explained at full length in Appendix A of
Ref. ~\cite{HTWI}, we will not repeat the details here. We choose 
$Q_0^2=5\,GeV^2, \Lambda=0.2\,GeV$ and, taking into account gluon to
quark conversion, the result at small $s$ is
\beqn
\beta q_i(\beta,Q^2)\simeq d\beta(1-\beta)(1-s)+
\zenn
\beq
+\varepsilon s\left\{\frac{4}{3}d\beta(1-\beta)\left(1+\ln\left(
\frac{(1-\beta)^2}{\beta}\right)\right)+\frac{1}{2}a\left(
\frac{2}{3}-\beta^2+\frac{\beta^3}{3}+\beta\ln\beta\right)\right\}
\label{z15}
\zen
with
\beqn
a=a(Q_0^2),\ d=d(Q_0^2)
\zenn
and
\beqn
\varepsilon=\frac{6}{33-2f}
\zenn
where $f$ is the number of quark flavors.
\vskip 0.3cm
With three flavors, $f=3$, the pomeron structure function is
\beq
G_{q/\pom}(\beta,Q^2)=\frac{4}{3}\beta q_i(\beta,Q^2).
\label{z16}
\zen
Parameters $a$ and $d$ are not independent since we impose the
momentum sum rule in the form
\beqn
\int_0^1 d\beta\,\beta\left[\sum_i q_i(\beta,Q_0^2)+G(\beta,Q_0^2)
\right]=1
\zenn
obtaining the constraint
\beq
d=\frac{1}{2}(2-a)
\label{z17}
\zen
While a proof of the validity of this sum rule for the pomeron
is lacking ~\cite{CHF}, its applicability appears reasonable once a
model for the pomeron in terms of its constituent is assumed 
~\cite{BERG,PRYTZ,GOUL}.
\vskip 0.3cm
Insisting on the particle nature of the pomeron we can evaluate
the total momentum carried by quarks and antiquarks
\beqn
M(Q^2)=6\int_0^1 \beta q_i(\beta,Q^2)d\beta=d+\frac{s}{9}
\left(a-\frac{77}{9}d\right)
\zenn
and, taking into account the relation (\ref{z17}), we find that
the $Q^2$ dependence disappear for $a=154/95$.
\vskip 0.3cm
This value is not far from the one obtained in fitting to the data,
\beqn
a=1.458,
\zenn
that gives a weak dependence on $Q^2$ with $dM(Q^2)/dQ^2<0$.
The relative contribution of quarks to the pomeron momentum is near 0.25 
in the range of $Q^2$ where the recursive method applies.
The presence of quarks at every $Q^2$ scale simulates a quarkball
~\cite{CF}.
\vskip 0.3cm
A plot of all ZEUS data for $F_2^{D(3)}$ ~\cite{ZEUS1} for different
$\beta$ values, regardless their $Q^2$ value, as in figures (2a-2c),
shows that the $Q^2$ dependence is indeed weak. The large errors 
permit only to say that our model is compatible with data.
Continuous and dashed lines are the result of the evolution at 
$Q^2=16$ and $28\,GeV^2$, respectively, with the nonlinear trajectory
(\ref{z8}) in the flux. No visible change is noticed if the calculation
is repeated with the linear trajectory (\ref{z7}).
\vskip 0.3cm
In comparison the H1 data, presented in Fig. (2d) for one $\beta$ value show
a larger spread in $Q^2$. In this case the selected values of $Q^2$ for the 
fit are $Q^2=12\, GeV^2$ (continuous curve) and $25\, GeV^2$ (dashed
curve)

\vskip 0.5cm
{\bf 5. Predictions and conclusions.}

In order to get a significant picture of the predictions from the
model, we integrate over $\beta$ and $Q^2$ the differential cross
section expressed as
\beq
\frac{d^4\sigma^{diff}}{d\beta dQ^2 d\xi dt}=\frac{2\pi\alpha^2}{\beta Q^4}
\left[ 1+(1-y)^2 \right] F_{\pom/p}(\xi,t)G_{q/\pom}(\beta,Q^2)~,
\label{z18}
\zen
where
\beqn
 y\simeq \frac{Q^2}{s\beta\xi}
\zenn
and $s=(296 \,GeV)^2$. The $\beta$ range of integration is
restricted to the interval
\beqn
 0.02\leq \beta \leq 0.8
\zenn
and the upper limit avoids the region near $\beta=1$ where
theoretical and experimental uncertainties are larger.
\vskip 0.3cm
The $Q^2$ integration has been performed for two different
intervals, both comprised in the region less sensitive to theoretical
approximations. The starting scale $Q^2=5\,GeV^2$ for evolution is
the lower limit in both cases and the $d^2\sigma/d\xi dt$
results are displayed in Figs. 3a, 3b. Predictions based on the 
logarithmic trajectory (\ref{z8}), continuous lines in Fig. 3,
are clearly distinguishable from the ones obtained with the linear
trajectory (\ref{z7}), dashed lines. Future experimental data,
even at moderate $|t|$ values, will be able to decide between
the two possibilities.
\vskip 0.3cm
    The $t$ dependence shown in Figs. 1, 3a, 3b has a nature
typically diffractive, i.e. it is strongly peaked in the forward direction.
At present, the (nearly) exponential decrease of the differential 
cross section $d^2\sigma/d\xi dt$ to large extent has been fed in by the 
choice of the residue and form of the pomeron trajectory.
Nevertherless, this result is far from being trivial since  in the formalism
under consideration, the $t$ dependence is correlated with other variables
and ultimately the choice of different inputs (pomeron trajectories)
will be tested experimentally.
\vskip 0.3cm
It is well known - both from the S-matrix theory ~\cite{BUG}
and from the fits to 
hadronic data - that the pomeron trajectory contains a significant 
non-linear part. The amount and form of this non-linear correction is a 
matter of debate in the literature. Our choice of the pomeron trajectory
(\ref{z8}) was based on earlier fits ~\cite{DJL}
to the data on high energy elastic hadron
scattering. Data on the $t$ dependence of the diffractive 
structure function have been discussed recently by two 
experimental groups: ZEUS ~\cite{SOL} and by UA8 ~\cite{SCL2}. 
We just notice that the UA8 results show the same trend in 
the behaviour of the differential cross section that derives from the
use of the logarithmic trajectory (\ref{z8}).
The results of the measurement of the $t$ dependence at HERA
are preliminary and still not complete.
\vskip 0.3cm
The knowledge of the $t$ distribution will settle important questions.
For example, the photon-pomeron vertex may be also 
$t$ dependent and it may bias the discussed behaviour of 
the structure function. 
\vskip 0.3cm
Another prominent feature in the $t$ dependence of diffraction, well known in
elastic hadron scattering, is the apperance of the diffraction minimum. 
The expected effect should be visible "by eye" 
because of its unmistakable structure, but the energy (here, $W$)
and $t$ values where the minimum should appear
are near the kinematical boundary of the relevant experiments on deep inelastic 
diffractive scattering. For the above (kinematical) reason it
has not yet been seen even in hadronic diffractive dissociation. Our 
model, as well as others, does not contain this structure. Its experimental 
observation however would resolve all doubts concerning the diffractive
nature of the "large rapidity gap events" and related phenomena.

\vskip 1.5cm
\underline {Acknowledgement}: We thank Prof. P.Schlein for showing us UA8 
results before publication and M. Arneodo, A. Solano for useful 
discussions on ZEUS results. One of us (L.L.J.) is grateful to the 
Dipartimento di Fisica dell'Universit\`a di Padova, to the Dipartimento di 
Fisica dell'Universit\`a della Calabria and to the Istituto Nazionale di 
Fisica Nucleare - Sezione di Padova and Gruppo collegato di Cosenza for 
their warm hospitality and financial support while part of this work was 
done.

\newpage
\centerline{\bf Figure Captions}
\vskip 0.3cm
\begin{description}
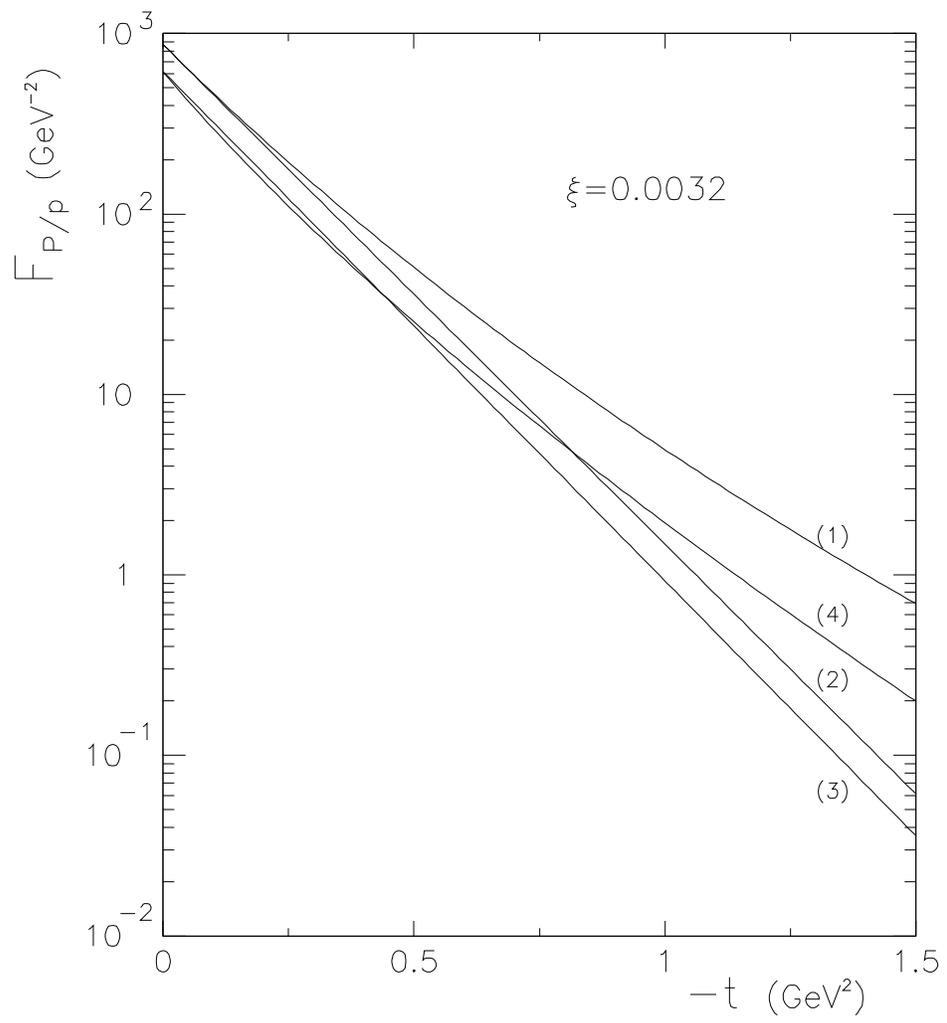
\item{Fig. 1:} The pomeron flux $F_{\pom/p}(\xi,t)$ versus $-t$
for $\xi=0.0032$ with the logarithmic trajectory (1) or the linear
one (2) discussed in the text. For comparison the pomeron fluxes
of Refs.~\cite{BERG}, (3) and ~\cite{DL}, (4) are also drawn.
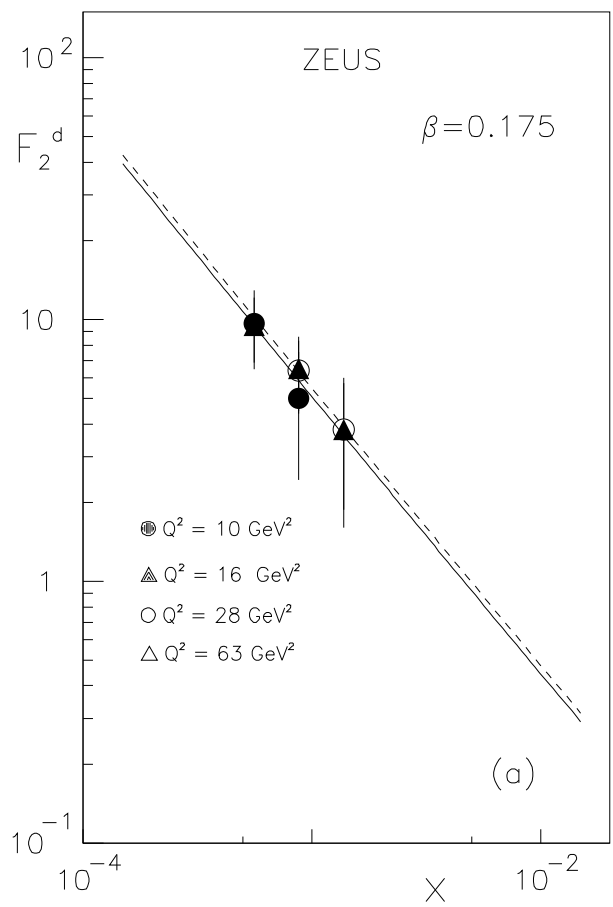
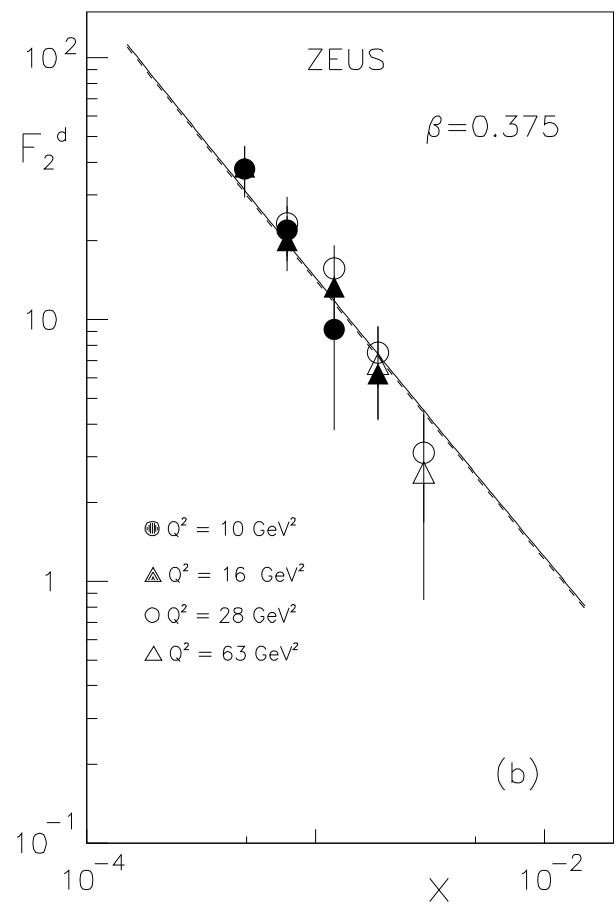
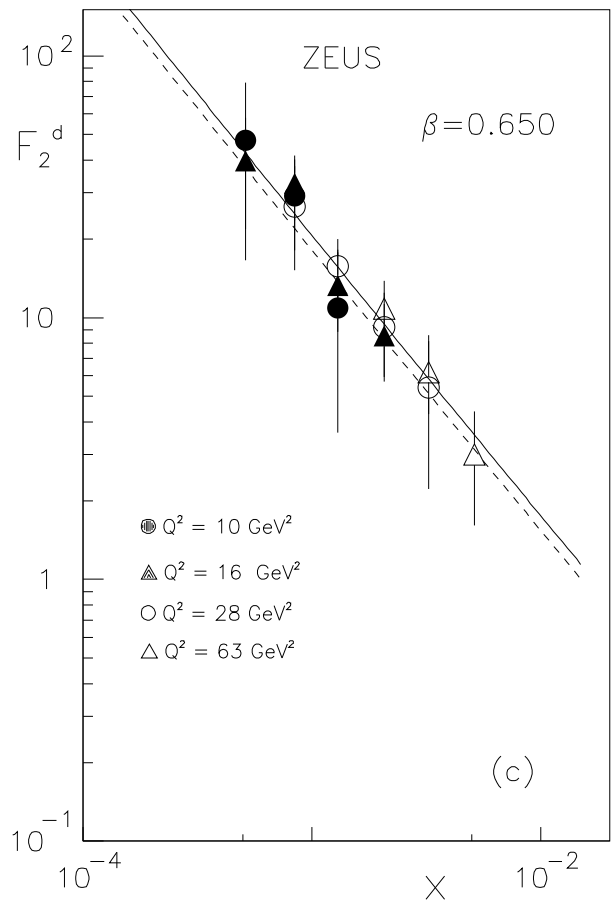
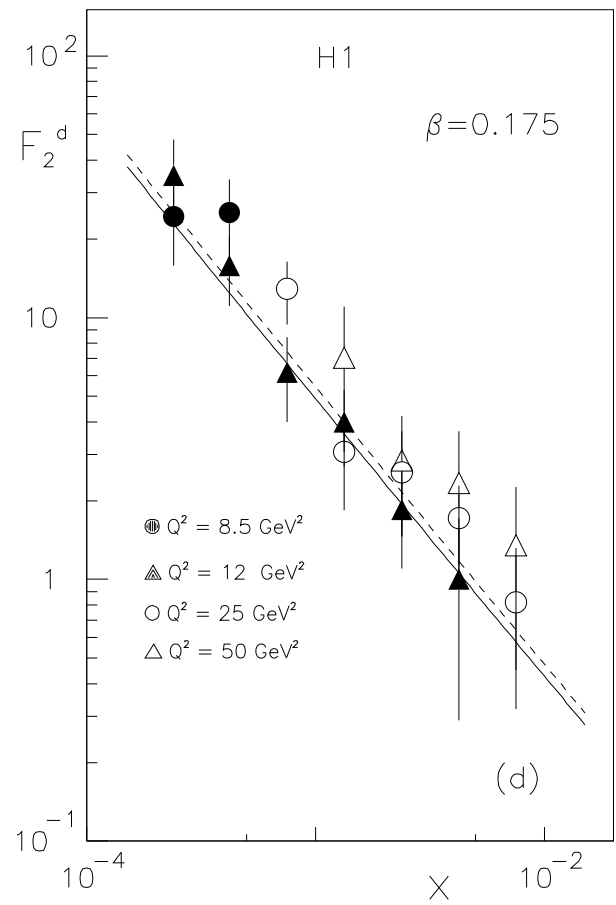
\item{Fig. 2:} The diffractive structure function $F_2^{D(3)}(\beta,
Q^2,x)$ versus $x$, for different values of $\beta$ and $Q^2$
compared with the ZEUS data ~\cite{ZEUS1} (a), (b), (c) and with the 
H1 data ~\cite{H1}, (d). $F_2^{D(3)}$ has been evolved at
$Q^2=16\,GeV^2$ (continuous line) and $Q^2=28\,GeV^2$ (dashed
line) for the ZEUS data. For the H1 data, $Q^2=12\,GeV^2$ (continuous
line) and $Q^2=25\,GeV^2$ (dashed line).
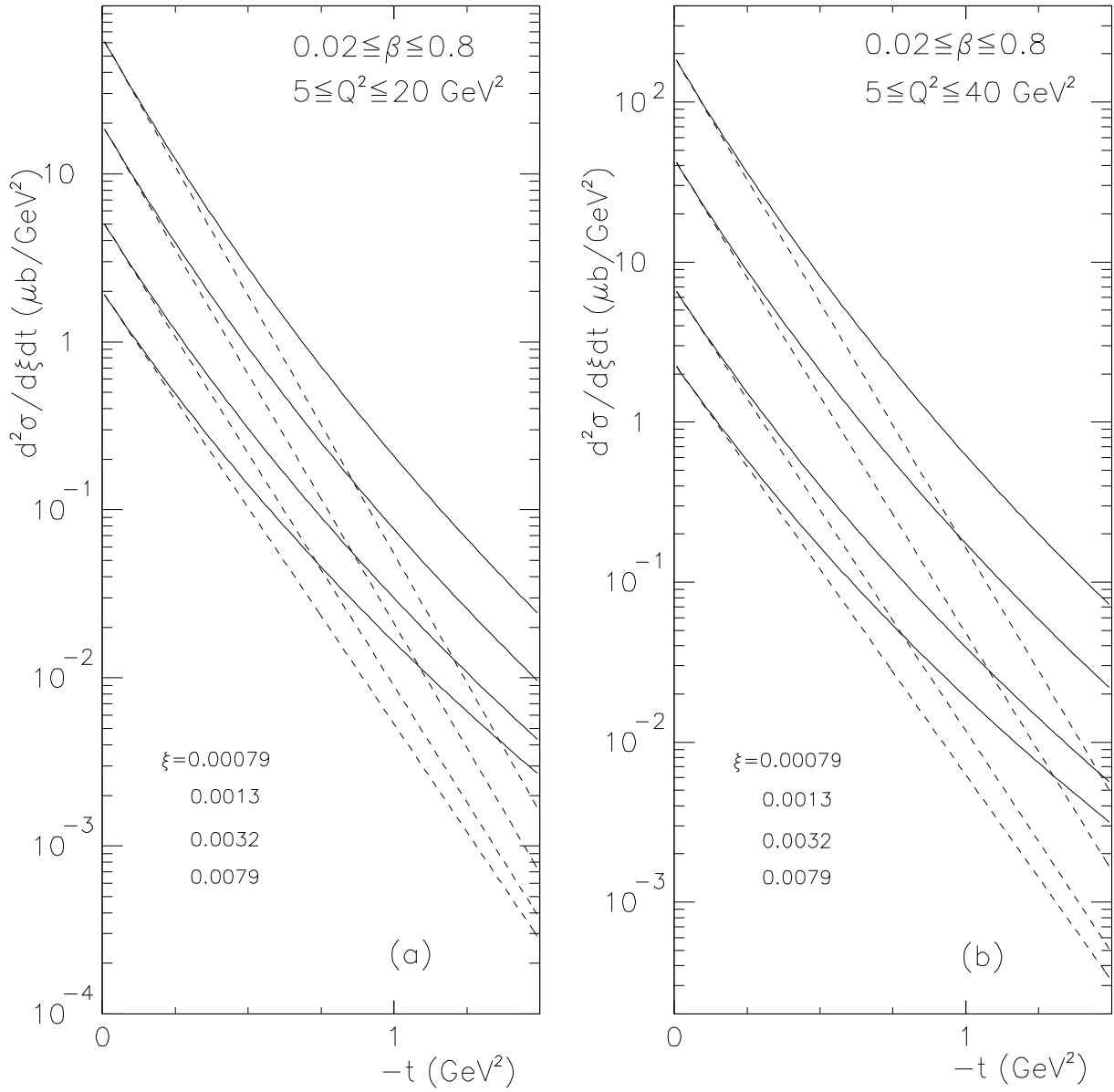
\item{Fig. 3:} The diffractive cross section $d^2\sigma/d\xi dt$
predicted from the model, versus $-t$ for different $\xi$ values.
The order of the $\xi$ values is respected in the curves. Two
different integration regions for $Q^2$ are considered: (a)
$5\,GeV^2\leq Q^2 \leq 20\,GeV^2$ and (b) $5\,GeV^2 \leq Q^2
\leq 40\,GeV^2$. Predictions from the logarithmic trajectory (full
lines) and from the linear one (dashed lines) are shown.

\end{description}

\end{document}